\documentclass[aps,showpacs,twocolumn]{revtex4}
\usepackage{psfig}
\usepackage{epsfig}
\usepackage{times}

\newcommand{\beq}{\begin{equation}}
\newcommand{\eeq}{\end{equation}}
\newcommand{\bea}{\begin{eqnarray}}
\newcommand{\eea}{\end{eqnarray}}
\newcommand{\nn}{\nonumber}

\newcommand{\veps}{\varepsilon}
\newcommand{\al}{\alpha}
\newcommand{\s}{\sigma}

\newcommand{\de}{\delta}

\newcommand{\be}{\beta}

\newcommand{\ua}{\uparrow}
\newcommand{\da}{\downarrow}

\newcommand{\dmi}{\frac{1}{2}}

\begin{document}

\title{Splitting electronic spins with a Kondo double dot device}
\author{Denis Feinberg,$^{a)}$ and Pascal Simon,$^{b)}$}
\affiliation{$^{a)}$ Laboratoire d'Etudes des Propri\'et\'es Electroniques
des Solides, CNRS, BP 166, 38042 Grenoble, France}
\affiliation{$^{b)}$ Laboratoire de Physique et Mod\'elisation des Milieux
     Condens\'es, CNRS et Universit\'e Joseph Fourier, 38042 Grenoble, France}
\date{\today}
\begin{abstract}
We present a simple de  made of two small  capacitively coupled quantum
dots in parallel.  This set-up can be used as
an efficient  "Stern-Gerlach" spin
filter, able to simultaneously produce, from a normal metallic lead,
two oppositely spin-polarized currents when submitted to a local
magnetic field. This proposal is based on the realization
of a Kondo effect where spin and orbital degrees of freedom are entangled,
allowing a spatial separation between the two spin polarized
currents. In the low
temperature Kondo regime, the efficiency is very high and the de 
conductance
reaches the unitary limit, $\frac{e^2}{h}$ per spin branch.
\end{abstract}


\maketitle

Controlling the electron spin in electronic circuits is the challenge of the
new emerging field called spintronics. Part of current research aims at 
the injection of spin-polarized electrons into mesoscopic structures.
For instance, control of single spins, owing to long
decoherence times in semiconductor nanostructures, opens the way to
spin-based quantum information processing. \cite{loss_book}
One of the major goals is the production of efficient spin
filters, with the following requirements : i) high polarization,
especially for very demanding tests of quantum entanglement \cite{Bell};
  ii) bidirectional spin filtering, e.g. filtering at will "up" and 
"down" spins;
iii) low impedance, to allow
unperturbed transport (conductance and noise) measurements on a variety of
de s.
Several set-ups fulfilling part, but not all the above constraints,
have been proposed or tested to inject spins or create spin
filters. They rely either on ferromagnetic materials \cite{ferro},
external magnetic fields \cite{recher,hanson,potok,borda},
or spin-orbit coupling. \cite{spin-orbit}
Recher {\it et al.} \cite{recher} have in particular considered a
quantum dot weakly coupled to current leads, in the sequential
tunneling regime. They have shown that in the presence of a local
magnetic field
it can act as an efficient spin filter whose spin direction can be
controlled by energy filtering, with the help of the dot
plunger gate voltage : given a single electron level in the dot, with
occupancy $n$, transitions between $n=0,1$ states or between $n=1,2$ states
respectively involve opposite spins. Another interesting possibility
developed by Borda {\it et
al.}\cite{borda} is to use a double quantum dot (DD) system with
strong capacitive interdot coupling.
When an external magnetic field is applied to such a system, these
authors showed
that the low energy physics can be described by a purely orbital
Kondo effect where spin flip
processes are suppressed and only charge fluctuations are allowed
between the dots (the latter representing the orbital degrees of
freedom). A major consequence of the Kondo effect
is the reach of the unitary limit at $T\ll T_K$ where $T_K$ is the
Kondo temperature.\cite{unitary}
In this limit the DD proposed by Borda {\it et al.}\cite{borda}
thereby acts as an almost perfect unidirectional spin filter (with
high conductance $e^2/h$), provided the temperature is low enough.

In this Letter, we go one step further and
propose a simple, robust and efficient  "Stern-Gerlach" spin
splitter, able to simultaneously produce from a normal metallic lead
two oppositely spin-polarized currents, using non-magnetic
semiconductor materials. Realization of such a spin splitter, used as
a source or an analyzer of polarized electrons, opens the way to many
experiments, including Bell correlations of entangled states,
\cite{Bell} or spin-resolved shot noise measurements. \cite{sauret}
Our proposal is schematized in Figure 1. Spin filtering is achieved by energy
filtering, as in Ref. \onlinecite{recher}, selecting each of the spin
directions
in either dot $1$ or $2$. Our
set-up does not
work in the sequential regime, but in the Kondo regime, as
   in Ref. \onlinecite{borda}, the two small quantum dots being strongly
coupled in a capacitive way. Rather than coupling each dot to two independent
reservoirs, here
each dot is connected to a common source kept at the chemical
potential $\mu_L$
    and to a distinct current lead at chemical potential $\mu_R$. The
two outgoing spin-polarized currents of opposite polarizations emerge
from these two separate leads.
\begin{figure}[h]
\epsfig{figure=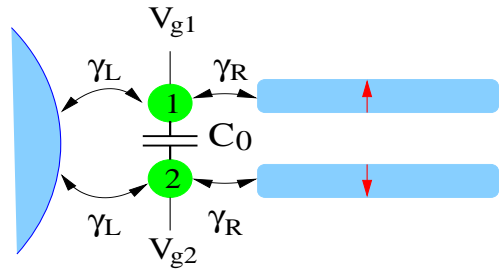,width=6.5cm,height=3.5cm}
\caption{Schematic representation of the proposed setup:
two small quantum dots coupled by a capacity $C_0$ and connected to a
common source.
Each dot $1$ and $2$ is also connected to an extra lead from 
which the two spin polarized
currents will emerge.
Depending on how the gate voltages are tuned, the upper lead can be
polarized in the up direction and the lower lead in the opposite
direction or  vice versa. }
\end{figure}

The numbers of electrons in the dots are controlled by two
plunger gate voltages at potential $V_{g1}$ and $V_{g2}$.
We label the lowest-lying charging states by the numbers $(n_1,n_2)$
of extra electron in dots $1,2$.
We consider the regime where
the gate voltages are adjusted such as
the two lowest-lying and almost degenerate charging states are $(1,1)$
and $(0,2)$, instead of states $(1,0)$ and $(0,1)$ as in Ref. [\onlinecite{borda}]. A schematic stability diagram is depicted in Fig. 2
showing the different possible charging states.
At energies lower than the charging energy of the dot
$E_C=min(E(0,1)-E(1,1);E(1,2)-E(1,1))$,
the charge dynamics is restricted to states $(1,1)$ and $(0,2)$,
states $(0,1)$ and $(1,2)$ appearing only as virtual states. Let us label the
capacitances (choosen symmetric in the dot indices for simplicity) as $C_{L}$
(left), $C_{R}$ (right), $C_{g}$, $C_0$ (coupling the two dots), define
$C=C_L + C_R + C_g$, and the external charges
$C_{g}V_{g1}$, $C_{g}V_{g2}$ from a reference state with even occupation
numbers. Then the intradot and interdot charging energies are respectively
$U=\frac{e^2(C+C_0)}{2C(C+2C_0)}$ and $V=xU$ with $x=\frac{C_0}{C+C_0}$.
The condition for degeneracy reads
$V_{g2}=V_{g1}+\frac{e}{C_g}$, and the excitation energies are
$E(0,1)-E(1,1)=U[(1+x)\frac{C_{g}V_{g1}}{e}-\frac{1}{2}]$,
$E(1,2)-E(1,1)=U[\frac{1}{2}+x-(1+x)\frac{C_{g}V_{g1}}{e}]$.
The optimum regime is reached at the symmetric
point O  (corresponding to the thick
black dot in Fig. 2) where the two excitation energies are equal to
$E_c=U\frac{x}{2}$.

\begin{figure}[h] 
\label{diagramm}
\epsfig{figure=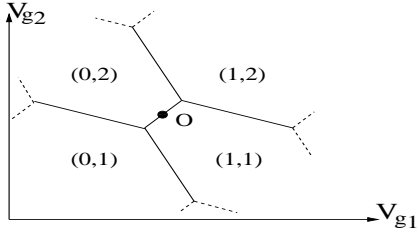,width=5.5cm,height=3.cm}
\caption{ Sketch of the operation region
in the stability diagram showing stable charge states as function of
gate voltages $V_{g1}$ and $V_{g2}$.
The thick black dot O corresponds to the ideal operating
point in the middle of the degeneracy line.}
\end{figure}

The isolated DD system is described\cite{borda} at low energy by
\beq \label{hdot} H_{dot}=-\delta E T^z-t T^x-g\mu_B B S^z,\eeq
where we have defined the orbital pseudospin $T^z=(n_1-n_2+1)/2=\pm
1/2$. Here $\delta E=E(0,2)-E(1,1)$ is zero when the two lowest-lying
charge states are
exactly degenerate. The second term in Eq. (\ref{hdot}) represents a
direct tunneling amplitude between the dots and the last term
expresses the Zeeman splitting when a local magnetic field is
applied in the z direction.

In the following we assume that the Zeeman energy is large enough
such that spin-flip scattering is suppressed. An evaluation of the
typical value of the required magnetic field and other experimental
parameters is provided at the end of the Letter. The spin
states of the two degenerate ground states
are therefore $(\ua,\ua)$ and $(0,s)$ where $s$ stands for singlet
state. Triplet states $(0,t)$ can be discarded if the level splitting
$\de \veps$ in each dot is large enough.
\cite{recher,hanson} Defining the total spin operator by
$S^z=S^z_1+S^z_2$, it is crucial that $T^z=S^z-\dmi$. This means that the
spin of the electron added to the ``empty'' state $(0,\ua)$ is entangled
with the orbital pseudo-spin:\cite{simon} virtual charge
fluctuations  on dot $1$ (resp. $2$) involve spin-up (resp.
spin-down) electrons exclusively, and an orbital pseudo-spin flip
(between states
$(\ua,\ua)$ and $(0,s)$) is
locked to a genuine spin flip. Therefore the Kondo screening of the
spin involves spin-up electrons in the upper right lead and spin-down
electrons in the lower right one, as well as spin-up and spin-down
electrons altogether in the common left lead.  This is contrary to the
set-up of Borda {\it et al.},\cite{borda} where the real spin is
quenched by the applied magnetic field and only the orbital
pseudo-spin survives. This is a crucial difference which makes
possible the realization of a spin splitter from our proposal.
Before turning to a more quantitative analysis, we also
emphasize that the tunneling term, being spin independent, no longer
connects the two degenerate states, as opposed to the
situation occurring in [\onlinecite{borda}]. We can therefore neglect
it provided $t<< g\mu_B B$. In practice, this makes a strong
capacitive coupling
between the dots easier to achieve than in Ref. \onlinecite{borda}
where $t << T_K$ is instead required.

The leads are described by
$H_{\rm leads}=\sum\limits_{k,\al,\s}\veps_k c^\dag_{k,\al,\s} c_{k,\al,\s}$,
where $ c^\dag_{k,\al,\s}$ creates an electron with energy
$\veps_k$ in the lead
$\al$ and spin $\s$.  Indeed, Zeeman splitting in the reservoirs can
be made much smaller
than in the dots. \cite{recher}
Since the Coulomb energy $E_C$ is one of the largest energy scales,
only cotunneling processes
where the numbers of initial and final electrons in the DD are equal
are allowed. Therefore we need only to consider virtual excitations
towards states with
$n_1+n_2=1$ and $3$ in the DD. Using
second-order perturbation theory in the tunneling amplitude between
the dots and the leads, we obtain
a Kondo effective Hamiltonian $H_{eff}=H_K+H_{tun}$ with
\bea \label{hkondo}
H_K&=&\sum\limits_{k,k',\al,\be}\left[J_{k,k',\al,\be}(c^\dag_{k,\al,\da}T^+c_{k',\be,\ua}+h.c.)
\right]\\
&+&\sum\limits_{k,k',\al,\be}\left[J_{k,k',\al,\be}
T^z(c^\dag_{k,\al,\ua}c_{k',\be,\ua}-c^\dag_{k,\al,\da}c_{k',\be,\da})\right]\nn
\eea
the Kondo part involving  the flip of the pseudo-spin $\vec T$ and
\beq \label{htun}
H_{tun}=\sum\limits_{k,k',\al,\be,\s}\left[
V_{k,k',\al,\be}(c^\dag_{k,\al,\s} c_{k',\be,\s}+h.c.)\right]
\eeq
corresponds to tunneling between the leads.
In these expressions the lead index $\al,\be$ takes only the two values $L,R$.
Due to spin-orbital entanglement, the distinction between the two
    right leads is simply given by the spin index: $(R,\ua)$ corresponds
to the upper right lead and $(R,\da)$ to the lower right lead. The
coupling $V$ being spin independent, there is no tunneling
between the two right leads.
As usual we can neglect the $k$ dependence of
the Kondo couplings i.e. $J_{k,k',\al,\be}\approx J_{\al,\be}\sim
\sqrt{\gamma_\al\gamma_\be}/E_C$ and $V_{k,k',\al,\be}\sim
J_{\al,\be}/4$, with $\gamma_\al$ the tunneling rate from/to lead $\al$.
There are other cotunneling terms with smaller amplitudes involving
for example higher energy processes like $E_C(2,1)-E_C(1,1)\gg E_C$.
These terms may a priori pollute spin filtering. Nevertheless the
strength of the Kondo effect is to renormalize the Kondo couplings
toward strong coupling
at low energy as opposed to direct potential scattering terms
that do not renormalize. Therefore terms like those involved in Eq.
(\ref{htun}) or other higher energy potential scattering terms can be
dropped out in the low temperature regime $T\ll
T_K=D\exp[-1/\rho_0(J_{LL}+J_{RR})]$,
where a constant density of states $\rho_0$ has been assumed in the leads.
This corresponds to the unitary limit \cite{unitary} where the spin 
and orbital pseudospin
are completely screened and a singlet is formed together with spin-up/down
electrons in the left lead, spin-up electrons in the upper right lead
and spin-down electrons in the lower right lead.

Transport across the double dot can now be described, applying a small
voltage $eV=\mu_L-\mu_R$. The conductance of each right lead is given
by $G_{L,R1,\ua}=G_0sin^2\delta_{\ua}$ and
$G_{L,R2,\da}=G_0sin^2\delta_{\da}$, both tending towards $G_0
=\frac{4\gamma_L\gamma_R}{(\gamma_L+\gamma_R)^2}\frac{e^2}{h}$ for
$T<<T_K$ where the phase shifts $\delta_{\ua}$, $\delta_{\da}$ are
equal to $\pi/2$ . The conductances reach $e^2/h$ at $T=0$ for symmetric
tunneling amplitudes. Notice that in the unitary limit, the
polarization of the currents in the right leads is almost perfect,
e.g. $G_{L,R1,\da}=G_{L,R2,\ua}\sim 0$.

The ground state in the unitary limit is a Fermi liquid which is
usually  stable toward
various perturbations. Let us analyze them in details. First, $\delta
E$ in (\ref{hdot}) plays the role of an orbital magnetic field
lifting the degeneracy between the dot $1$ and $2$ levels. Therefore
the two plunger gate voltages $V_{g1}$ and $V_{g2}$ need to be finely
tuned such that $\delta E\ll T_K$, in order to reach the unitary
limit. In addition, we have assumed from the beginning that the
tunneling amplitudes
between the dots and their respective right leads are equal. Actually,
   a different tunneling amplitude would lift the spin degeneracy
between electrons
with spin up and down, playing a role similar to an effective
magnetic field. The situation is also analogous to that of a single
   quantum dot in the Kondo regime, with ferromagnetic reservoirs
breaking the symmetry between up and down spins. In this situation,
the spin dissymetry can be compensated by applying a magnetic field.
\cite{martinek,choi} In the present case, one would simply need
to correct by slightly modifying the gate voltages $V_{g1}$ or $V_{g2}$
(except \cite{choi} at the symmetric point O).

Let us now estimate the experimental requirements to
realize our proposal. First, in a large enough magnetic field,
the above set-up should exactly map on a single-dot Kondo problem.
This requires,
as in Ref. [\onlinecite{borda}], that $g\mu_BB > T_K^{(0)}$, which is the Kondo
temperature of the system without a magnetic field whose behavior is 
also expected to
be described at low energy by a SU(4) Kondo problem with four 
(orbital and spin) degenerate states.
\cite{borda,simon} Note that due to the higher symmetry $T_K^{(0)}\gg T_K$.
Then, the main conditions
are set by the necessary Zeeman splitting $g\mu_BB$, e. g. $g\mu_BB <
\de \veps$, to
avoid populating a triplet state in dot $2$, more precisely
$\de \veps-g\mu_BB > T_K$. The working conditions can then be summarized as
$T < T_K < T_K^{(0)} < g\mu_BB < \de \veps, E_C$, and $t < g\mu_BB$.
They are consistent with small dots with large magnetic fields, e. g. typically
$T_K \sim 0.1K$,  $T_K^{(0)} \sim 0.5K$, $B\sim10T$ with $g=0.44$,
$\de \veps\sim U\sim V \sim 1meV$.
These conditions are comparable to the ones proposed
in ref. [\onlinecite{borda}] and an efficiency of around $95\%$
can be reached at low enough temperature $T\ll T_K$.

As a conclusion, we have used an exotic Kondo effect, where
spin and orbital degrees of freedom are entangled, to propose a
simple, robust and efficient
spin splitter. 
We hope that this proposal will open new
opportunities in the field of spintronics based on the application
of the Kondo effect in semiconductor quantum dots.

\end{document}